\documentstyle{mn}

\input epsf

\begin{document}

\newcommand{\COBE}{{\sl COBE}}
\newcommand{\kms}{\,{\rm km}\,{\rm s}^{-1}}
\newcommand{\hMpc}{\,h^{-1}{\rm Mpc}}
\newcommand{\Ob}{\Omega_{{\rm B}}}

\journal{Preprint SUSSEX-AST 96/5-2, UBC-COS-96-01, astro-ph/9605057}
\title[CDM models with high baryon content]{Cold dark matter models with 
high baryon content}
\author[M. White et al.]{Martin White,$^1$ Pedro T. P. Viana,$^2$
Andrew R. Liddle$^2$ and Douglas Scott$^3$ \\
$^1$Enrico Fermi Institute, University of Chicago, 5640 S.~Ellis Ave, 
Chicago, Illinois 60637,~~~U.~S.~A.\\
$^2$Astronomy Centre, University of Sussex, Falmer, Brighton BN1 9QH, U. 
K.\\
$^3$Department of Physics \& Astronomy, 129-2219 Main Mall, University of 
British Columbia, Vancouver, B.~C.~V6T 1Z4,~~~Canada}
\pubyear{1996}
\maketitle

\begin{abstract}
Recent results have suggested that the density of baryons in the
Universe, $\Ob$, is much more uncertain than previously thought, and
may be significantly higher. We demonstrate that a higher $\Ob$
increases the viability of critical-density cold dark matter (CDM)
models.  High baryon fraction offers the twin benefits of boosting the
first peak in the microwave anisotropy power spectrum and of
suppressing short-scale power in the matter power spectrum. These
enable viable CDM models to have a larger Hubble constant
than otherwise possible. We carry out a general exploration of high
$\Ob$ CDM models, varying the Hubble constant $h$ and the spectral
index $n$. We confront a variety of observational constraints and
discuss specific predictions. Although some observational evidence may
favour baryon fractions as high as 20 per cent, we find that values
around 10 to 15 per cent provide a reasonable fit to a wide range of
data. We suggest that models with $\Ob$ in this range, with
$h\simeq0.5$ and $n\simeq0.8$, are currently the best critical-density
CDM models.
\end{abstract}


\section{Introduction}

There is presently greater uncertainty regarding the baryon density of
the Universe than there has been for many years. The standard theory
of big bang nucleosynthesis (BBN), developed over more than 40 years
and applied extensively to local abundance data, has long been
considered as fixing the baryon density to a satisfyingly high level
of accuracy.  A typical quoted value a few years ago was $\Ob
h^2=0.0125\pm0.0025$, where the uncertainty is to be interpreted as
something like 95 per cent confidence \cite{Waletal,Smietal}.  Here
$\Ob$ is the baryon density in units of the critical density, and $h$
is the present Hubble parameter in units of 100$\kms\,{\rm Mpc}^{-1}$.
However, recently it has been acknowledged that there are a range of
possible systematic uncertainties in determinations of the primordial
abundances \cite{Wiletal,SasGol,Skietal,Scuetal}, which have led to a
broadening of the preferred interval, particularly towards the upper
end.  Recent quoted ranges, at 95 per cent confidence, include
$0.010<\Ob h^2<0.022$ by Copi, Schramm \& Turner (1995a,b), with a
similar result by Turner et al. \shortcite{TTSC}, and $0.0125<\Ob
h^2<0.0275$ by Hata et al. (1996).  Kernan \& Sarkar
\shortcite{KerSar} quote a 95 per cent confidence upper limit $\Ob
h^2<0.032$, while an analysis by Krauss \& Kernan \shortcite{KraKer}
has also suggested higher baryon densities, although without quoting a
specific range.

Added to that, it is now becoming possible to measure element
abundances at high redshift as well as locally, through absorption
systems in the spectra of distant quasars.  Results for deuterium have
so far been inconclusive (Songaila et al. 1994; Carswell et al. 1994,
1996; Wampler et al. 1996; Rugers \& Hogan 1996a,b; Tytler, Fan \&
Burles 1996; Burles \& Tytler 1996), with some very high abundances
reported (implying a low baryon density) and also some low ones. There
may be some reason to favour the lower determinations, in that most of
the likely systematic effects, such as interloper hydrogen clouds,
bias the abundance estimates upwards.  However, only determinations in
a large number of separate systems will truly resolve the issue. The
most recent determination giving a low deuterium abundance
\cite{TytFanBur,BurTyt} yields the result
$\Ob h^2=0.024\pm0.002\pm0.002\pm0.001$, with the 1$\sigma$
uncertainties being statistical, systematic, and theoretical
respectively.  This is still consistent with the more recent determinations
using local abundances, while reinforcing the view that the baryon
density may be higher than previously thought.

Within the conventional picture, the new deuterium results can only be
made consistent with BBN if the primordial helium abundance $Y_{\rm
P}$ has been underestimated; $\Omega_{\rm B}h^2=0.025$--0.035
corresponds to 25 per cent helium, rather than the usual 23--24 per
cent.  However, this seems to be entirely within the bounds of
possibility (e.g.~Olive \& Steigman 1995; Sasselov \& Goldwirth 1995;
Burles \& Tytler 1996), given possible systematic effects.  The low
deuterium values also imply very little destruction of primordial
deuterium, in order to be consistent with the interstellar medium
values, which give a constraint around $\Omega_{\rm B}h^2\le0.031$
\cite{McC,Linetal}.  Hence values of $\Omega_{\rm B}\ga0.15$ (assuming
$h\simeq0.5$) begin to run into quite firm limits.  However, it may
also be possible to relax the nucleosynthesis bound on $\Ob$ with new
particle physics, an example being the decaying tau neutrino proposal
of Gyuk \& Turner \shortcite{GT}, or by allowing inhomogeneities in
the baryon-to-photon ratio (see e.g.~Mathews, Kajino \& Orito 1996 and
references therein).

The other salient observational issue is the question of the baryon
fraction in clusters of galaxies
\cite{Whietal,WhiFab,ElbArnBoh,Maretal}, which consistently gives
values which are high for a critical-density universe.  For example, a
recent compilation by White \& Fabian \shortcite{WhiFab} gives
\begin{equation}
{\Ob\over\Omega_0} = 0.14_{-0.04}^{+0.08} \, \left(h\over0.5\right)^{-3/2}
\end{equation}
again at the 95 per cent level.  There does seem to be variation in
this quantity for individual clusters, but it is still unclear whether
this is due to systematic effects.  Certainly some clusters appear to
have a baryon fraction as high as 20 per cent (see e.g.~Mushotzky
1995).  At the moment there is no obviously reliable lower limit to
the cluster baryon fraction, although the values adopted by Steigman
\& Felten (1995), ${\Ob/\Omega_0}\ge 0.2 \left(h/0.5\right)^{-3/2}$,
and by Evrard, Metzler \& Navarro (1996), ${\Ob/\Omega_0}\ge 0.11
\left(h/0.5\right)^{-3/2}$, are typical of the range.  Assuming there
has been no significant segregation of the baryons, and that there has
been no serious systematic underestimation of the mass of galaxy
clusters (see e.g.~Gunn \& Thomas 1995; Balland \& Blanchard 1996),
this is hard to reconcile with a critical-density universe, unless
$\Ob$ has a higher value than given by the canonical BBN numbers.

There are some additional considerations which may favour an increase in the
baryon fraction, although they would not make strong arguments on their
own.  One example is that a higher value of $\Ob$ may bring simulations of
Lyman alpha absorption systems more in line with data \cite{HKWM,KWHM,MCOR}.
Another example is the apparently high baryonic mass fraction in the halo of
our Galaxy, in the form of MACHOs \cite{Alcetal,MACHO2}.  And a third is
that higher $\Ob$ may help proto-galaxy disks be more self-gravitating,
thereby alleviating a problem with over-concentration of baryons
in simulations of galaxy formation (see e.g.~Steinmetz 1996).

We are interested in the implications of these changing notions about
$\Ob$ for models of structure formation, and in particular for the
cold dark matter (CDM) model \cite{Pee82,Bluetal,Davetal} in the case of
a critical-density universe (see Davis et al. 1992a and Dodelson, Gates
\& Turner 1996 for recent reviews).  We will argue that a factor of
two increase in the predicted baryon content from its circa 1993 value
has a far from insignificant effect on the viability of such models,
and indeed that there exist models from this class which provide a
reasonable fit to the current data.  The bottom line will be that
models which are selected to conform with a wide range of constraints
have: (1) moderately tilted initial conditions; (b) $h\simeq0.5$; and
(c) $\Ob$ centred around a value of say 12 per cent.

The outline of the paper is as follows. In the next section we give an
overview of the situation regarding critical-density CDM models, and
in the following Section define our terminology and motivate our
choice of model parameters.  We then compare these high $\Ob$ models with 
the available data on structure formation and on the cosmic microwave
background (CMB) in Sections 4 and 5 respectively.  We conclude
with a discussion of some of the strengths and weaknesses of these models
and future key observations.

\section{The big picture}

In order to obtain a critical-density universe of sufficient age, it
is necessary to choose a Hubble parameter not much greater than
$h=0.5$ (for example, an age greater than 12 Gyr requires $h<0.55$).
Similar values for $h$ are also necessary if one is to obtain a
satisfactory structure formation model \cite{WSSD,LLSSV}.  Some recent
observational analyses have lent support to such a low Hubble constant
\cite{Sch,Tametal,Braetal}, and such values even remain marginally
consistent with the Freedman et al.~\shortcite{Frietal} measurement of
$h=0.80\pm0.17$. With $h$ around 0.5, the predicted age of the
universe appears to be consistent with the ages of the oldest globular
clusters (Bolte \& Hogan 1995; Chaboyer et al. 1995, 1996; Jiminez et
al. 1996).  For this value of $h$, the old nucleosynthesis
determinations would indicate a universe with about 5 per cent
baryons, while the higher values favoured more recently would be more
like 10 per cent. As we shall see, this increase is far from
insignificant, and indeed could be regarded as a new `standard'
value\footnote{In fact, many $N$-body simulations and other CDM
calculations of the 1980s adopted this round number as a working
value, and it is still sometimes used in this context.}.  The cluster
number, centred around 15 per cent and going up to above 20 per cent,
suggests the possibility of an even higher value.

If one is determined to retain a critical-density universe, it is
extremely interesting that there is now strong evidence, from a
structure formation perspective, in favour of these higher $\Ob$
values.  The demonstration of why this makes such a difference is the
principal goal of the present paper.  The gist of the argument is as
follows. It is well known that, despite the strong appeal of
simplicity and physical motivation, the standard CDM model (sCDM),
based on a scale-invariant initial spectrum ($n=1$), a Hubble constant
$h=0.5$ and with 5 per cent baryons, is unable to fit all the
observational data.  This is largely because, when normalized to
reproduce the microwave anisotropies detected by the Cosmic Background
Explorer (\COBE) satellite \cite{Ben96}, the power spectrum has too
much short-scale power.  If one is to retain the CDM hypothesis,
modifications are required, and two have received particular
attention, the first being to reduce the Hubble parameter \cite{BBST}
and the second to tilt the spectrum of density perturbations
\cite{VML,Bond,LLS,CGKO,MMdGV,PolSta}. The former has the drawback
that it appears necessary to reduce $h$ to about 0.35 \cite{LLSSV},
which is lower than most researchers are willing to accept. The latter
has the problem that if one tilts the spectrum sufficiently to remove
the unwanted short-scale power in the matter spectrum, then the tilt
is also enough to remove the peak from the CMB anisotropy spectrum
\cite{WSSD,Whi}.

A range of CMB data now implies the existence of a power-spectrum peak
at sub-degree scales (Scott, Silk \& White 1995; Kogut \& Hinshaw
1996), which seems even more compelling with the recent Saskatoon-95
(SK95) observations \cite{Netetal}. This means that purely tilting the
spectrum, while retaining the other standard CDM parameters, is
beginning to look untenable. The situation improves somewhat if one
combines a smaller tilt with a more modest reduction in $h$
\cite{WSSD,LLSSV}, but as we shall see this is apparently still not
enough if the currently favoured peak height is confirmed by future
experiments.

The most natural way to try to resolve this conflict, while staying
within the CDM paradigm, is to raise the baryon fraction. This has
twin benefits.  Regarding the radiation power spectrum, the effect of
the extra baryons is to enhance the first peak. With enough baryons,
this can go a long way towards compensating for the loss of height
introduced by the tilt (the tilt in turn being required if one is to
prevent the required $h$ being too small), and hence retain
compatibility with the recent CMB observations.  The second benefit is
that the baryons themselves help suppress the short-scale power in the
matter power spectrum, since unlike the CDM they are unable to
collapse until after decoupling. This means that the correct power
spectrum shape can be obtained for a larger Hubble parameter, or a
more modest tilt (cf.~White et al.~1995b).

Before embarking on our discussion of critical-density CDM models, let
us comment on some of the alternatives.  One can reduce the
short-scale power by assuming that a component of the dark matter is
hot \cite{ShaS,BV,DSS,TR,KHPR,Jinetal}.  At present this is a
perfectly satisfactory solution (especially if a modest tilt is also
allowed \cite{LidLyt,SchSha,PogSta,LLSSV}), but loses the simplicity
of having only cold dark matter. Even more fanciful solutions exist,
including decaying neutrinos \cite{BE91,DGT,WGS,McNP}, broken power
laws or double inflation \cite{TVVSJ,Gotetal,PetPolSta,Katetal}, and
an additional isocurvature component \cite{Kawetal,SBG}.  There is
also the possibility that astrophysical processes could play a role in
determining the power spectrum of galaxy fluctuations
\cite{BabWhi,Efs,Bowetal,Lametal}, although it is not at all clear
that such effects will be significant.

A more fundamental change is to stay with CDM but abandon
critical-density models, going instead to low density. Viable models
of this type are possible both in open universes
\cite{RP,GRSB,LLRV,YamBun} and in flat universes with a cosmological
constant \cite{Pee84,TSK,KS,ESM,SugSut,EBW,Kofetal,KT,OS,SGB,LLVW}.
This has been seen as the easiest way to make CDM models consistent
with the observational constraints, as indicated by the large number
of papers exploring this possibility.  We note though that pressure
has been exerted on both the open and flat cosmological constant dominated
approaches recently.  The combination of the SK95 \cite{Netetal} and
CAT \cite{Scoetal} CMB observations appears to fix the location of the
acoustic peak at around $\ell\simeq 200$, suggesting a flat geometry
rather than a hyperbolic one.  Furthermore, the somewhat lower
normalization of the four-year \COBE\ data
\cite{Ben96,Ban96,Goretal,Hin96,Wrietal} exacerbates the problem that 
low-density open universes have a low normalization.  Meanwhile,
measures of the deceleration parameter using Type Ia supernovae
\cite{Peretal} suggest a universe which is decelerating rather than
accelerating, which for a flat universe requires $\Omega_0>2/3$,
agreeing with the quasar lensing constraint \cite{Koc}.  In addition,
types of observations which have historically favoured high density,
such as velocity power spectrum measures \cite{Dekel,StrWil,Zaretal},
redshift-space distortions (Cole, Fisher \& Weinberg 1995; Taylor \&
Hamilton 1996), the properties of voids \cite{NusDek,DekRee}, and the
absence of a prominent break in the non-linear power spectrum
\cite{KlyPriHol}, continue to do so.

So, although there are many alternative paths to explore, it seems to
us premature to relinquish the aesthetic appeal of critical density
CDM models.  Within the context of this general paradigm, there remain
a number of parameters to be honed by the data.  The value of the
baryon fraction was traditionally a fixed quantity, and also thought
to be low enough that the cosmological effects would be relatively
unimportant.  This is no longer true if one allows for the higher
values of $\Ob$ which now seem more credible, and moreover such
variations are now significant at the level of detail which current
data can distinguish.  Raising $\Ob$ to 10 per cent or above makes a
substantial difference, as we will now show.

\begin{figure*}
\begin{center}
\leavevmode \epsfysize=10cm \epsfbox{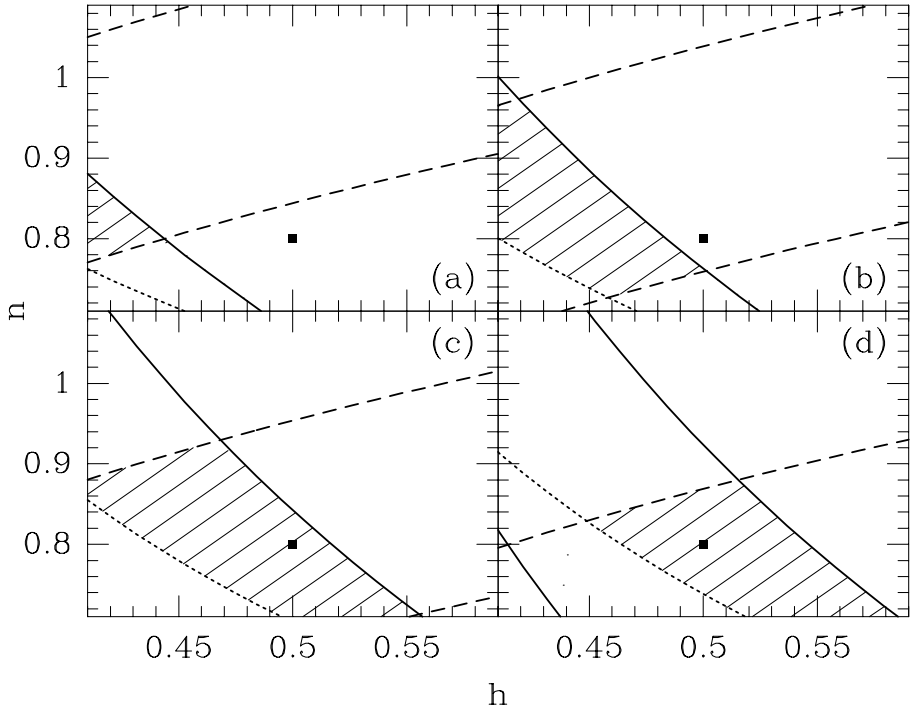}
\end{center}
\caption{Linear theory constraint curves for CDM models in the
$h$--$n$ plane for different values of $\Ob h^2$, namely (a) $0.0125$,
(b) $0.025$, (c) $0.0375$ and (d) $0.05$.  The squares indicate where
the models in Table 1 lie.  The curves and shaded areas show 95 per
cent confidence regions from the galaxy correlation function (solid),
the height of the CMB peak (dashed) and the abundance of damped Lyman
alpha systems (dotted).  It is the tension between these constraints
which is eased by allowing a higher $\Ob$.  Cluster abundance, quasar
abundance and velocity information are less constraining than these.
The shaded region is allowed by all data constraints, and some
continuation above the dashed line would be permitted if reionization
was sufficiently early.}
\label{fig:linh}
\end{figure*}

\section{The Dark Matter Power Spectrum}

The matter power spectrum in the CDM model is specified by two
functions.  The initial spectrum of perturbations is taken to be a
power-law $P_{{\rm init}}(k)\propto k^n$, where $n$ is the slope.  The
case $n=1$ is the scale-invariant spectrum, and $n \neq 1$ models are
often referred to as `tilted'.  Inflationary models typically give $n$
around one, though the possible range covers all the region of
interest for structure formation\footnote{Inflationary models may also
produce a spectrum of gravitational waves which can influence the {\sl
COBE} normalization; typically these make it more difficult to fit the
data and we do not consider them here.}.  The transfer function $T(k)$
measures the amount of growth (relative to the infinite wavelength
mode, and computed in linear theory) experienced by the modes of
wavenumber $k$.  The present linear power spectrum is given by
$P(k)\propto P_{{\rm init}}(k)\,T^2(k)$.  In $\Omega=1$ CDM models, the
transfer function is time-independent at late times; however, its form
depends on the cosmological parameters $h$ and $\Ob$ describing the
matter content of the Universe.  Commonly, CDM models are described
via a fit to the standard form\footnote{This form is not a good
approximation at high $\Ob/\Omega_0$ unless $h\simeq0.50$.  This could
be important when considering low $\Omega_0$ models should $\Ob$ turn
out to be 10 per cent or higher \cite{BDP}.} of Bardeen et al.
\shortcite{BBKS}
\begin{eqnarray}
T(q) & = & \frac{\ln \left(1+2.34q \right)}{2.34q} \times
        \\ \nonumber
& & \hspace*{-0.5cm}
        \left[1+3.89q+(16.1q)^2+(5.46q)^3+(6.71q)^4\right]^{-1/4} \,,
\end{eqnarray}
with $q = k/h\Gamma$ \cite{Sug,HuSug}. The parameter $\Gamma$, a function of
both $h$ and $\Ob$ and referred to as the `shape parameter', describes
the shape of the matter {\it transfer function}.  We have calculated
transfer functions for our models numerically and then fit to this
form.

In observational papers, the best-fitting model for a particular data
set is often given in terms of an effective shape parameter,
$\Gamma_{{\rm eff}}$.  This is usually done by fitting to the power
spectrum under the assumption that $n=1$.  Since we will be allowing
$n$ different from unity, and also since the fit to the shape will
depend on the range of scales considered, there is some ambiguity in
the meaning of this $\Gamma_{\rm eff}$, and so we will be explicit in
our definition here. The `effective' value of $\Gamma$ for the
theoretical models which we show in the table is obtained by finding
the scale-invariant $\Gamma$-model which gives the same ratio of power
on scales $8\hMpc$ and $50\hMpc$, those scales being chosen as roughly
the range typically probed by galaxy surveys.  Since fitting a
scale-invariant $\Gamma$-model to a tilted model is only approximate,
it must be treated as only the roughest of guides and we shall always
work from the true power spectrum.

Following Liddle \& Lyth \shortcite{LLRep}, the normalization can be
specified by considering the value of the density fluctuation at
horizon-crossing: $\delta_{{\rm H}}^2(k) \propto k^{n-1}$. This is
defined through the contribution to the rms fluctuation per
logarithmic interval in wavenumber $k$ as
\begin{equation}
\Delta^2(k) \equiv {k^3 P(k)\over 2\pi^2} \equiv
  \left( {k\over H_0} \right)^{4}\ \delta_{\rm H}^2(k)\ T^2(k) \,,
\end{equation}
which is dimensionless. The normalization can be specified by giving
the value of $\delta_{{\rm H}}$ at the present Hubble scale. The
four-year \COBE\ normalization of tilted critical-density CDM models
can be well represented by a fitting function
\begin{eqnarray}
\delta_{\rm H} (k = H_0) & = & \left( 1.94\pm 0.13\right) 
	\times 10^{-5} \\ \nonumber
& & \exp\left[ -0.95(n-1)-0.169(n-1)^2 \right],
\end{eqnarray}
which is accurate to better than 3 per cent for $0.7<n<1.2$ \cite{BunWhi}.
The effects of $\Omega_{\rm B}$ and $h$ on this fit are negligible.

A quantity related to the power spectrum is the variance of the density
field smoothed on a scale $R$, which is simply the integral of $\Delta^2$,
\begin{equation}
\sigma^2(R) = \int {{\rm d}k\over k}\ \Delta^2(k)\ W^2(kR) \,,
\end{equation}
where we take the smoothing function $W(kR)$ to be a spherical
top-hat, given by
\begin{equation}
W(kR) = 3 \left( {\sin(kR)\over (kR)^3}-{\cos(kR)\over(kR)^2} \right) .
\end{equation}

Ways of comparing models with large-scale structure data are well
developed in the literature. We use observations as described in
Liddle et al. (1996b,c) to outline the preferred regions of parameter
space (see also Silk 1994; Bond 1994; Bahcall 1995; White et al.
1995b; Primack 1996).  Four crucial types of data turn out to be the
most constraining, namely: the \COBE\ observations of large-angle
anisotropies; a compilation of intermediate-scale microwave anisotropy
data; the abundance of damped Lyman alpha systems seen in quasar
absorption spectra; and the shape of the galaxy correlation function.
We shall consider several other types of data in addition to these,
which prove not to give firm additional constraints.

\begin{figure}
\begin{center}
\leavevmode \epsfysize=6cm \epsfbox{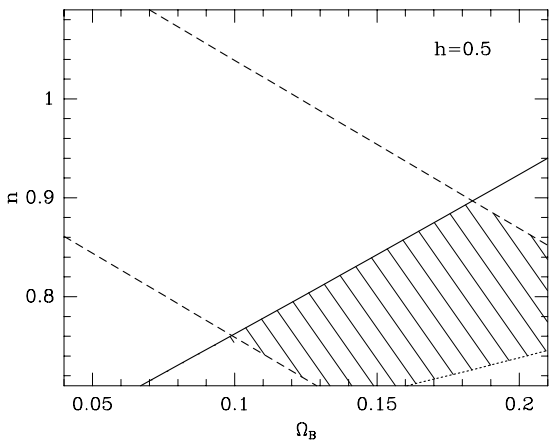}
\end{center}
\caption{As Fig.~1, but in the $\Ob$--$n$ plane with $h$ fixed at 0.50.
The small region to the bottom right is ruled out because it has
insufficient power to produce damped Lyman alpha systems at $z=4$.
The continuation of the shaded region upwards across the dashed line
would be permitted by early reionization.}
\label{fig:linb}
\end{figure}

We shall describe all of these in more detail in the following
sections.  However, to guide our later discussion, we show the results
of the linear theory comparison now.  The three parameters we can vary
are $\Ob$, $h$ and $n$.  Fig.~\ref{fig:linh} shows cuts of the
parameter space at different choices of constant $\Ob h^2$. In each
case there is an allowed region, but for the old nucleosynthesis value
of $\Ob$, shown in Fig.~\ref{fig:linh}(a), agreement is available only
for very low $h$.  There is considerable pressure from direct
measurement against reducing $h$ below 0.5, and it is rapidly
problematic for large-scale structure to go much above it, whatever
the baryon density.  For this reason we also show, in
Fig.~\ref{fig:linb}, a parameter-space cut at the value $h=0.5$.  This
highlights the importance of varying the baryon density, and in many
ways this figure can be said to represent our main result.  Only by
raising the baryon density to 10 per cent and above can
critical-density CDM models be viable at $h=0.5$. Both the galaxy
correlation function and the CMB peak height exert pressure in that
direction.

In order to focus our discussion, we select some particular models for
closer study. We choose to concentrate on $h=0.5$, for which $n=0.8$
is around the optimal choice, although less tilt can be accommodated
for lower values of $h$.  Our main aim is to consider variations of
$\Ob$, and so we pick four values, corresponding to: the old BBN
number (5 per cent); a number consistent with the low deuterium
measurements and more recent BBN calculations (10 per cent); the
favoured number from the cluster baryon fraction (15 per cent), which
is some way above even the highest standard nucleosynthesis upper limits;
and an extreme example representing close to the top of the range of
observed cluster baryon fractions (20 per cent).  In
Table~\ref{tab:someoldnumbers} we show a summary of important
quantities for models spanning this range, calculated primarily in
linear theory by numerical evolution of the coupled fluid, Einstein
and Boltzmann equations\footnote{In White et al. \shortcite{WSSD} the
values for $\sigma_1$ were obtained from extrapolating $P(k)$ as a
power law from $8\hMpc$ down to $1\hMpc$. This neglects the curvature
of the transfer function over this range of scales and over-estimates
$\sigma_1$.  We use the full transfer function here.}.  Here the
quantity $z_{{\rm ri}}$ is the redshift of reionization; it is
estimated using the techniques of Liddle \& Lyth (1995), and is highly
uncertain.  The optical depth to Thomson scattering,
\begin{equation}
\tau=0.035\,\Ob\,h\, \left[ (1+z_{\rm ri})^{3/2}-1 \right] \,,
\end{equation}
inherits this uncertainty, and measures the probability of microwave
photons being re-scattered, assuming full ionization from $z_{\rm ri}$
to the present.

\begin{table}
\begin{center}
\begin{tabular}{lcccc}
& \multicolumn{4}{c}{$\Ob$} \\
          &    $0.05$   &    $0.10$   &     $0.15$  & $0.20$ \\ \hline
$\Gamma$          &     0.44    &     0.41    &     0.39    & 0.36  \\
$\Gamma_{\rm eff}$&     0.36    &     0.33    &     0.31    & 0.28 \\
$\sigma_8$        &$0.77\pm0.06$&$0.72\pm0.05$&$0.66\pm0.05$&$0.61\pm0.05$\\
$\sigma_1$        &$2.8 \pm0.2 $&$2.5 \pm0.2 $&$2.3 \pm0.2 $&$2.0 \pm0.2 $\\
$\sigma_{0.2}$    &$5.0 \pm0.5 $&$4.5 \pm0.3 $&$3.9 \pm0.3 $&$3.4 \pm0.3 $\\
$z_{\rm ri}$      & 14          & 12          &      11     & 9 \\
$\tau$            & 0.05        & 0.08        &      0.10   & 0.11
\end{tabular}
\end{center}
\caption{Power spectrum measures for selected CDM models.  The models have
$\Omega_0=1$, $h=0.5$, $n=0.8$, with four values of $\Ob$ as shown.
The values of $\Gamma$ come from fitting to numerical evaluations of
the transfer function for $10^{-3}\le k\le 1\,h\,{\rm Mpc}^{-1}$.  The
values of $\Gamma_{{\rm eff}}$ are fits to the ratio of the
dispersions $\sigma_R$ at $R=50h^{-1}$ and $R=8h^{-1}$Mpc.  The errors
on $\sigma_R$ reflect the $1\sigma$ uncertainty from the {\sl COBE}
four-year normalization.  The estimates of the reionization redshift
$z_{\rm ri}$ and optical depth $\tau$ are highly uncertain.}
\label{tab:someoldnumbers}
\end{table}

\section{Structure Formation}

We now discuss the main limits imposed on cosmological models from
observations of the clustering properties of galaxies, the galaxy
cluster abundance, peculiar velocities and high redshift object
formation.

\begin{figure}
\begin{center}
\leavevmode
\epsfysize=6cm \epsfbox{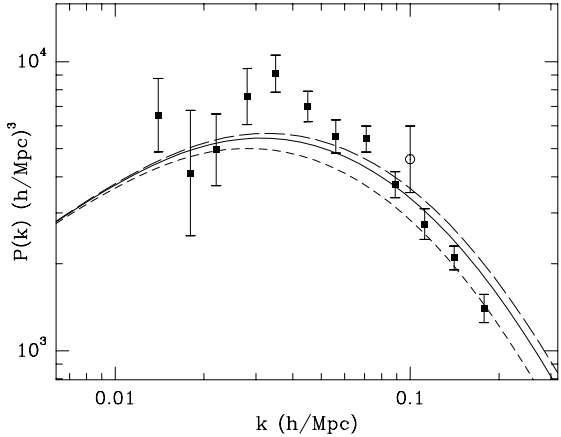}
\end{center}
\caption{The matter power spectrum for $\Ob=0.05,0.1,0.2$, the highest
lines having the lowest baryon density (we omit the 15 per cent model
for the sake of clarity).  All models have $h=0.5$ and $n=0.8$.  The
solid squares are taken from Peacock \& Dodds (1994) with the 4 points
at highest $k$ omitted.  We have adjusted the data to the best fit
normalization for the $\Ob=0.1$ model.  The open circle is the
velocity-derived point from Kolatt \& Dekel (1996); note however that
the error bar shown on this point omits cosmic variance, which is
large.}
\label{fig:pk_pd}
\end{figure}

\subsection{The galaxy correlation function}

Perhaps the classic problem of standard CDM has been that the model
has the wrong `shape'.  Specifically the ratio of large-scale
($\sim50\hMpc$) power to small-scale ($\sim8\hMpc$) power is too small
compared to observations.  Both tilt and small-scale damping from a
high baryon fraction help address this.

The current situation regarding the data relating to shape is somewhat
uncertain.  If we compare our models to the compilation of Peacock \& Dodds
\shortcite{PeaDod}, taking the error bars at face value, then we require
$\Ob\ge0.1$ to provide sufficient large-scale power, as can be seen in 
Fig.~3.  For the model
with $\Ob=0.1$ and $n=0.8$, a simple $\chi^2$ fit to the Peacock \&
Dodds \shortcite{PeaDod} data points assuming Gaussian uncorrelated
error bars and allowing the normalization to vary, shows the model is
excluded at about 97 per cent confidence, if one drops the four
shortest-scale points which may be affected by non-linear biasing
\cite{Pea}.  If all the points are included this model fits at the
$95$ per cent confidence level.  The models with higher $\Ob$ fare
better, being allowed at better than the 95 per cent confidence level,
even excluding the four shortest-scale points.  A reasonable
conclusion based on this constraint then is that critical-density CDM
models (with $h\simeq0.5$ and $n\simeq0.8$) are allowed provided that
$\Ob$ is about 10 per cent or greater.

\begin{figure}
\begin{center}
\leavevmode
\epsfysize=6cm \epsfbox{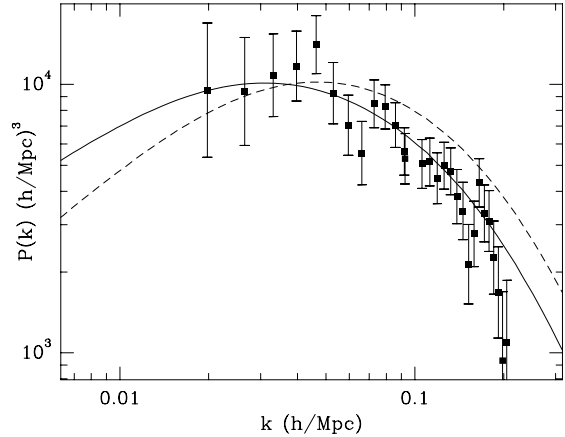}
\end{center}
\caption{The redshift space matter power spectrum for the $\Ob=0.1$
model from Table~1 (solid line).  The dashed line shows the result for
standard CDM. The data are taken, without adjustment, from Tadros \&
Efstathiou (1995; weight $8000$); approximately every second point is
uncorrelated. This ignores a downward correction to the theoretical
curves of roughly 15 per cent for $k<0.1\,h\,{\rm Mpc}^{-1}$, arising
from the window function of the surveys.}
\label{fig:pk_rs}
\end{figure}

The Peacock \& Dodds \shortcite{PeaDod} data are a compilation to
which several adjustments and/or corrections have been made. Since we
have specific models in mind, we can carry the predictions forward to
confront individual data-sets directly, rather than comparing to processed
data.  An example is shown in Fig.~\ref{fig:pk_rs}, where we compare the
$\Ob=0.1$ model to the power spectrum of {\it IRAS} galaxies in
redshift space, conveniently tabulated in Tadros \& Efstathiou
\shortcite{TadEfs}.  We have chosen this $\Ob$ as the model which is
only marginally allowed by Peacock \& Dodds \shortcite{PeaDod}.  In
this model the {\it IRAS} galaxies are almost unbiased tracers of the
mass -- we have not scaled the theory or the data points in
Fig.~\ref{fig:pk_rs} in any way.  Specifically, we have assumed a
redshift-space distortion parameter $\beta=1$, and a small-scale
exponential velocity distribution with $\sigma=280\kms$ to convert
from real to redshift space \cite{ColFisWei}.  It is clear from
Fig.~\ref{fig:pk_rs} that this model fares extremely well against this
survey, though it is less constraining than that of Peacock \& Dodds
\shortcite{PeaDod} since it incorporates less data.

\begin{figure}
\begin{center}
\leavevmode
\epsfysize=6cm \epsfbox{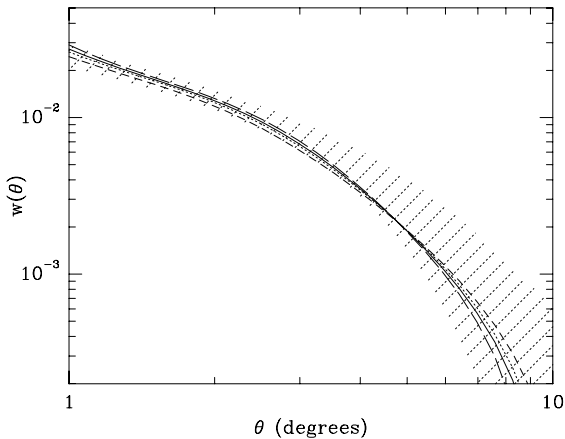}
\end{center}
\caption{The galaxy angular correlation function $w(\theta)$ for the APM
survey as predicted by our models with $h=0.5$, $n=0.8$ and $\Ob=0.05$
(long dashed), 0.10 (solid), 0.15 (dotted) and 0.20 (short dashed).
All curves have been \COBE\ normalized with bias chosen to agree with the 
data at $\sim1^{\circ}$.  They include a correction from the non-linear part 
of the power spectrum.  The shaded region represents the $2\sigma$ error 
band from a fit to the APM data (Baugh \& Efstathiou 1994).}
\label{fig:wtheta}
\end{figure}

The first firm indication that standard CDM lacked power on large
scales came from the APM survey \cite{Madetal,MadEfsSut}. To compare
to this survey, we show in Fig.~\ref{fig:wtheta} the predicted angular
correlation function $w(\theta)$ for our four models. We use the fit
in Jain, Mo \& White \shortcite{JaiMoWhi} to obtain the non-linear
power spectrum; this is particularly important on scales less than
$1\degr$, but also has a modest effect at larger angles where it
suppresses the correlation slightly. All our models clearly have
sufficient large-scale power to fit this data, in contrast to the
standard CDM model.

\subsection{Galaxy cluster abundance}

The present abundance of galaxy clusters requires $\sigma_8\simeq0.6$,
with most authors agreeing fairly well on the central value but with
differing opinions as to the uncertainty
\cite{E89,HA,BahCen,WEF,BMc,VL,ECF}. Many determinations do not permit
values as high as our lowest-baryon model predicts (see
Table~\ref{tab:someoldnumbers}), though Viana \& Liddle
\shortcite{VL} quote +32 per cent and -24 per cent at 95 per cent
confidence, among the most conservative, which does just allow it.
However, things are much better for the models with a higher baryon
content.  We find that any model of the type considered in this paper
satisfying both the {\sl COBE} and the galaxy correlation data
constraints automatically satisfies the cluster constraint as well.

\subsection{Peculiar velocities}

The galaxy peculiar velocity field can be recovered through the POTENT
procedure \cite{BD89} from peculiar velocity catalogs, such as the
Mark III catalogue.  It can then be used to constrain the amplitude of
the power spectrum either directly, through the spatial derivatives of
the peculiar velocity field \cite{Dekel,StrWil}, or indirectly, by
means of the bulk velocity \cite{KolDek}.  Because the latter is
sensitive to larger scales, results obtained through both methods do
not necessarily coincide.  Due to the very large cosmic variance,
resulting from the fact that only local measurements have been made
from a random field, we find that neither method further constrains
the available parameter space left by the other observations.
However, for comparison, we show in Fig.~\ref{fig:pk_pd} a point from
the POTENT velocity power spectrum of Kolatt \& Dekel
\shortcite{KolDek}.  Notice that our models fit this well even without
adding the cosmic variance.  We note in passing that the velocity data
can also be used to constrain the shape of the power spectrum, though
at the moment results are not as robust as those using galaxy
correlation data.  Zaroubi et al.  \shortcite{Zaretal} seem to prefer
a higher $\Gamma_{{\rm eff}}$, between 0.35 and 0.55 when marginalized
over $\sigma_8$, than the galaxy surveys, though the uncertainty is
large.

In addition, the question of the small-scale ($\sim1\hMpc$) velocities
should be addressed, i.e.~is the velocity field cold enough?  This
problem has been discussed by many people (see White et al. 1995b for
references) but a clean comparison with the data remains difficult.
Following White et al. \shortcite{WSSD}, we compute the second moment
of the mass correlation function $J_2(1\hMpc)$, which is related to
the rms velocity on the same scale.  We find a reduction in
short-scale power, as measured by $J_2(1\hMpc)$, by a factor of 4 to 5
for our $\Ob\ga0.1$ models, compared with standard CDM.  Although this
is a significant reduction, it may still be too hot relative to that
measured in redshift catalogues, though this may be subject to
substantial cosmic variance.  We believe that the situation regarding
velocities on these scales will probably require careful comparison of
$N$-body simulations with observational results, which is beyond the
scope of this paper.

\subsection{High-redshift object abundance}

We are going to concentrate on three types of objects found at high
redshifts: damped Lyman alpha system (DLAS), galaxies and quasars. 
We will also mention the situation regarding clusters at intermediate 
redshifts. The technique we use to compare observed abundances with the 
ones predicted by our models is the standard Press--Schechter
\shortcite{PreSch} calculation (see Liddle at al. 1996b for a
discussion of our particular implementation).

\subsubsection{Damped Lyman alpha systems}

The observations of DLAS by Storrie-Lombardi et al. \shortcite{Stoetal} 
imply that at redshift 4, the most constraining point, the amount of
neutral hydrogen in bound objects is
\begin{equation}
\Omega_{{\rm HI}}=(0.0011 \pm 0.0002)\,h^{-1} \,.
\end{equation}
To obtain a conservative bound, we assume that all the hydrogen
in DLAS is in the neutral state, and taking the $2\sigma$ lower limit, we
obtain a lower bound on the baryon density at that redshift,
\begin{equation}
\Omega_{{\rm B}}(z=4) \geq 0.0007\,h^{-1} \,.
\end{equation}
In order to compare this limit with our models we need to make some
further assumptions regarding the nature of the DLAS. To be
conservative we assume that the DLAS at such a high redshift arise
predominantly within filamentary structures, which seems to be
supported by recent hydrodynamical simulations \cite{KWHM}, and
consequently we use the threshold value of $\delta_{c}=1.5$ in the
Press--Schechter calculation \cite{Mo}.  We also conservatively use
the 95 per cent upper limit on the {\sl COBE} normalization, though
the models also fit well at the central normalization.

\begin{figure}
\begin{center}
\leavevmode \epsfysize=6cm \epsfbox{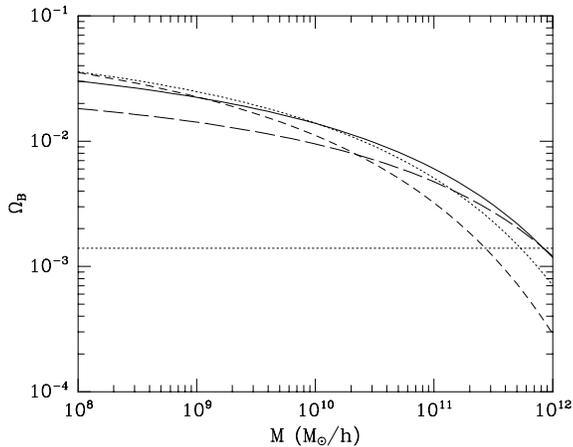}
\end{center}
\caption{The {\em baryon} density in objects of mass greater than $M$ at
redshift 4, for our models. Along the right-hand side, the higher lines
correspond to lower baryon density, the lines being 5 (long dashed),
10 (solid), 15 (dotted),and 20 (short dashed) per cent.
The horizontal dotted line indicates the 95 per cent lower limit from DLAS
observations (Storrie-Lombardi et al.~1995).}
\label{fig:highz1}
\end{figure}

In Fig.~\ref{fig:highz1} we show an estimate of the {\em baryon} density in
objects with a mass in excess of $M$ at redshift 4 for our models.
There is little difference at high redshifts between the models with higher
baryon fractions.  These models would come under pressure were it shown that
the observed systems correspond to substantially more massive objects than
usually supposed. However, under the reasonable assumption that the DLAS
correspond to objects of masses $10^{10}$ $h^{-1}$ ${{\rm M}}_{\sun}$ and
greater, all of the models are comfortably allowed.
We show the full parameter space constraints as the dotted curves in
Figs.~\ref{fig:linh} and \ref{fig:linb}.  The DLAS provide a limit on how 
much
tilt can be introduced; at high baryon density this is presently a stronger
constraint than the CMB peak height, as seen in Fig.~\ref{fig:linb}.

\subsubsection{Lyman break galaxies}

\begin{figure}
\begin{center}
\leavevmode \epsfysize=6cm \epsfbox{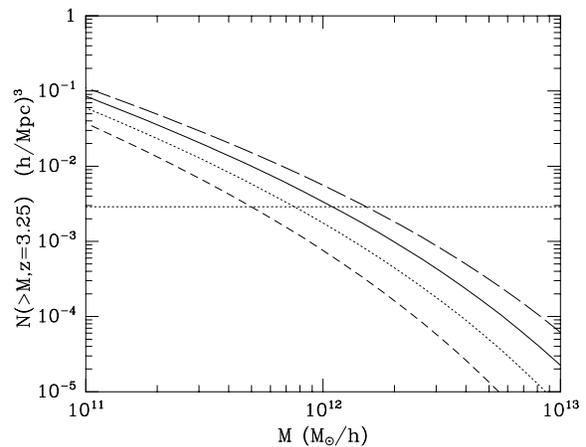}
\end{center}
\caption{The number density of collapsed objects with mass above 
$M$ for the four models at redshift 3.25. Higher lines correspond to
lower baryon density, the lines being 5, 10, 15 and 20 per cent. The
dotted line indicates the observed number density of Lyman break
galaxies (Steidel et al. 1996a).}
\label{fig:highz2}
\end{figure}

Recently Steidel and collaborators (Steidel at al. 1996a,b;
Giavalisco, Steidel \& Macchetto 1996) were able to detect galaxies between
redshifts 3 and 3.5 using a technique based on the identification of the
redshifted Lyman continuum break.  Their results put a lower bound on the
number density of Lyman break galaxies in that redshift range:
\begin{equation}
N_{{\rm gal}}(z=3.25) \geq 0.00288\,h^{3}\,{\rm Mpc}^{-3} \,.
\end{equation}
In Fig.~\ref{fig:highz2} we plot the number density of collapsed
objects with mass above $M$ at redshift 3.25, for our four specific
models. Here we use the threshold value of $\delta_{c}=1.6$ in the
Press--Schechter calculation, which seems the optimal when the objects
under consideration might have collapsed predominantly non-spherically
due to local tidal fields (Monaco 1995; Bond \& Myers 1996a,b). As
before, we conservatively use the 95 per cent upper limit on the
\COBE\ normalization.

In order to constrain our models we need a rough estimate of the mass
of the virialised halos with which the Lyman break galaxies are
associated.  According to the observations (Steidel et al. 1996b,
Giavalisco et al. 1996) the luminosity of these galaxies arises
predominantly within an approximately spherical central region with a
radius of about 2.5 $h^{-1}$ kpc, where the velocity dispersion is
between 180 to 320 $\kms$.  Assuming that galaxy halos can be roughly
approximated by a truncated singular isothermal sphere, these velocity
dispersions correspond to halo masses in the region of
$5\times10^{11}\, h^{-1} \, {{\rm M}}_{\sun}$ to
$3\times10^{12}\, h^{-1} \, {{\rm M}}_{\sun}$.
However, the quoted interval for the velocity
dispersion was obtained assuming that the observed broadening of
interstellar absorption lines is solely due to gravitational induced
motions. If part of the broadening is due to non-gravitational
effects, such as interstellar shocks arising from local star-forming
regions, then the velocity dispersions would be smaller and hence also
the derived halo masses.  On the other hand, the true mean velocity
dispersion of the halo is probably higher than the velocity dispersion
of the stars, as these feel the effect of only the central part of the
halo mass. In the future, spectra with higher signal to noise ratios and 
better resolution may enable the distinction between different sources of
line broadening.

Another problem for our models may be the observation by Steidel et al.
\shortcite{Steetala}, using additional {\sl K}-band photometry, that 
the Lyman break galaxies appear redder than one would expect if
they were undergoing their first burst of star formation.  However,
there remains considerable uncertainty in estimating formation time
from colours.  Nevertheless, were it shown that these galaxies formed
substantially earlier than the redshift at which they are observed,
all models with critical density would be under severe strain.

\subsubsection{Quasars}

For a long time, the abundance of quasars was the only information
available about the process of structure formation at high redshifts.
Now the abundances of DLAS, and more recently of Lyman break galaxies,
probe similar epochs. Further, it has been shown that in general both
DLAS abundances (see e.g.~Liddle et al. 1996b) and Lyman break galaxies
\cite{MF} provide stronger constraints on theoretical models of
structure formation than quasar abundances. The observed quasar number
density at a redshift of 3.25 is about $10^{-6}$ $h^{3}$ ${{\rm
Mpc}}^{-3}$ (Schmidt, Schneider \& Gunn 1995), and as can be seen in
Fig.~\ref{fig:highz2} this constraint is much weaker than that arising
from the abundance of Lyman break galaxies; all models are viable for
any quasar host galaxy mass up to $10^{13}$ $h^{-1}$ ${{\rm
M}}_{\sun}$. At even higher redshifts the observed quasar number
density does become more constraining, but not by much. For example,
at redshift 4 our models are still able to reproduce the observed
quasar number density, about $3\times10^{-7}h^{3}{\rm Mpc}^{-3}$
\cite{SSG}, for any host galaxy mass up to $10^{13}$ $h^{-1}$ ${{\rm
M}}_{\sun}$, except the model with $\Ob=0.2$ which requires host
galaxy masses from $6\times10^{12}$ $h^{-1}$ ${{\rm M}}_{\sun}$
upwards.

We also carried out a more careful analysis along the lines of Nusser
\& Silk \shortcite{NS}, allowing for relations between the abundance
of quasars and their mean lifetime, and the luminosity and mean
lifetime of a quasar and its minimum mass. We found that the
conclusions given above hold for quasar mean lifetimes over a broad
interval between about $10^{7}$ and $10^{9}$ yrs, the precise range
depending on the fraction of collapsed halos hosting quasars, the
amount of baryons in the halo used to form the central black hole and
the quasar radiative efficiency. Our models and conclusions are
similar to the tilted cases considered by Haehnelt (1993) and by
Nusser \& Silk (1993), who found that $n\simeq0.8$ models were
allowed, although the quasars were then required to be radiating near
to the Eddington limit and with lifetimes of $\sim10^7$ years.  Note
that for quasars, in common with other high-$z$ objects, there is a
great degree of uncertainty in the estimation of the amount of
collapsed matter at early times (see e.g.~Katz et al.~1994; Eisenstein
\& Loeb~1995), and that it may be easier to form central black holes
in higher $\Ob$ models.

\subsubsection{Clusters}

Finally, regarding intermediate redshift X-ray clusters, the best
available observations at this moment are those of Luppino \& Gioia
\shortcite{LG}.  Their sample of 6 clusters was selected from the {\it
Einstein} {\it Observatory} Extended Medium Sensitivity Survey (EMSS).
They have luminosities in excess of $L_{\rm X} = 1.25 \times 10^{44}
h^{-2}$ ergs ${{\rm s}}^{-1}$ and redshifts between 0.55 and 0.83,
where the most distant ever detected X-ray selected cluster is
located. Given the comoving volume over which the survey was carried
out, they conclude that the comoving number density of X-ray clusters
with luminosities in excess of $L_{\rm X}$ is $(8.8 \pm 3.44) \times
10^{-8} h^{3}{{\rm Mpc}}^{-3}$, at a mean redshift of 0.66.
  
Again the difficulty is in obtaining a reliable mass estimate for
these objects.  However if we instead concentrate on the cluster X-ray
temperatures, we can get a rough idea of what to expect from our
models at $z=0.66$.  For that we first need to relate the X-ray
luminosity $L_{\rm X}$ to an X-ray temperature.  To a first
approximation we can assume that the $L-T$ relation at zero redshift
also applies at $z=0.66$.  Following Henry et al. \shortcite{HGMMSW},
we translate $L_{\rm X}$ into a bolometric luminosity using $L_{{\rm
bol}}=1.18(kT)^{0.35}L_{\rm X}$, where $kT$ is in keV.  Using the
observed zero redshift $L-T$ relation given in David et al.~(1993), and
assuming $h\simeq0.5$, we find that $L_{\rm X}$ corresponds to an
X-ray temperature of roughly $6\,$keV.  Following Viana \& Liddle
\shortcite{VL}, and using the values of $\sigma_{8}$ in
Table~\ref{tab:someoldnumbers}, we calculate the mass corresponding to
these clusters under the constraint that their observed zero redshift
number density be recovered \cite{HA}.  We assume $\delta_{{\rm
c}}=1.6$ in the Press-Schechter calculation, integrate over
temperature and obtain an estimate of the number density of X-ray
clusters with $L_{{\rm bol}} > 1.1 \times 10^{45} h^{-2}$ergs ${{\rm
s}}^{-1}$ at $z=0.66$.  For our models with $\Ob=0.05$ and $\Ob=0.20$,
we obtain $4\times10^{\pm1}$ and
$2\times10^{\pm1}\times10^{-8}h^{3}{{\rm Mpc}}^{-3}$ respectively,
with the other models lying in between.  Here the uncertainty, an
order of magnitude in each direction, comes from pushing both the {\sl
COBE} normalization and the observed present number density of
$6\,$keV X-ray clusters \cite{HA} to their $2\sigma$ confidence
limits.  As can be seen, all our models have mild difficulty in
producing the observed number density of high-redshift clusters, but
in view of the uncertainties there is still plenty of room for
manoeuvre.  However, current data may suggest that at fixed
temperature the X-ray cluster luminosity is smaller at higher
redshifts (e.g.~Navarro, Frenk \& White 1995); if this trend turns out
to be correct, then {\it all} models with critical density will have
great difficulty in reproducing the observed number density of
high-redshift clusters.

\begin{figure*}
\begin{center}
\leavevmode
\epsfysize=11cm \epsfbox{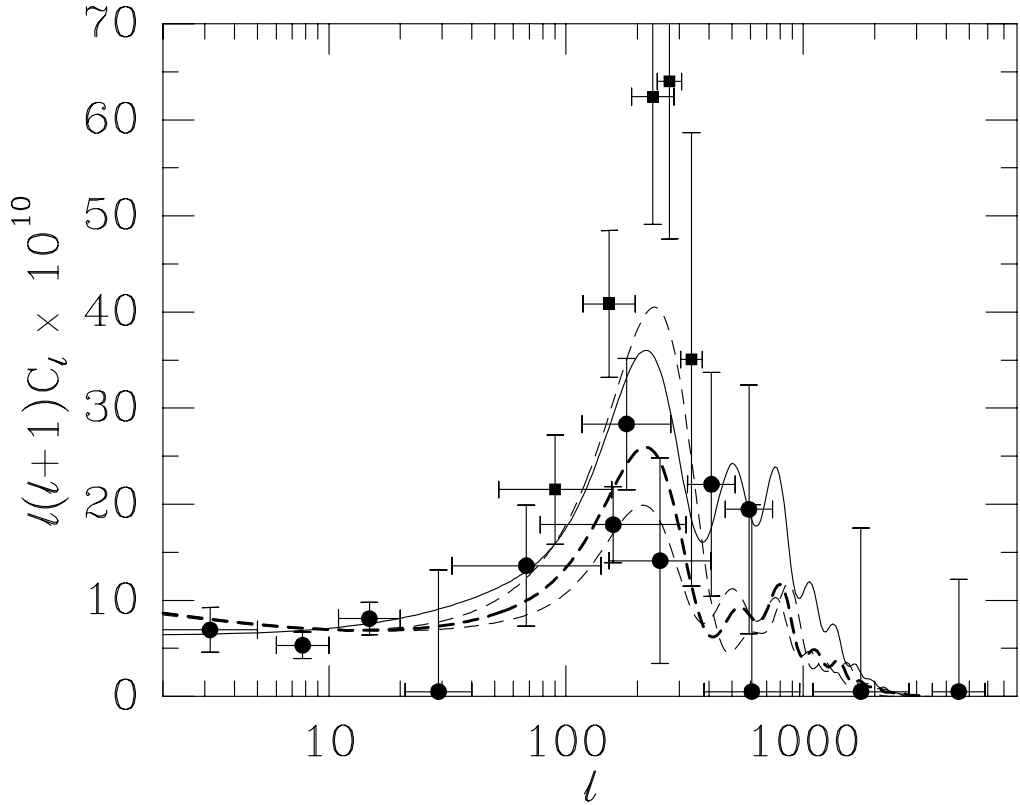}
\end{center}
\caption{The radiation power spectrum for our models (dashed) with
$\Omega_0=1$, $h=0.5$ and $n=0.8$. The curves have $\Ob=0.05$, 0.1
(thick) and 0.2 from bottom to top at $\ell\sim200$ (we have omitted
the $\Ob=0.15$ line for clarity).  The solid line is sCDM, with $n=1$,
$h=0.5$ and 5 per cent baryons.  All models have been normalized to
the 4-year {\sl COBE} data.  The points are recent 1$\sigma$
detections and 95 per cent upper limits as discussed in the text.
Solid circles from left to right are: {\sl COBE} (3 points, 1 upper
limit), SP94, MAX, Python, MSAM, CAT (2 points), White Dish, OVRO,
ATCA.  The solid squares are the SK95 points, without inclusion of the
calibration uncertainty in the error bars.  The error bars shown on
this figure should be regarded as indicative only.}
\label{fig:cl}
\end{figure*}

\section{Cosmic Microwave Background Anisotropies}

Any theory which purports to explain structure formation in the
universe must fit both the large-scale structure data and the data on
CMB anisotropies.  The most influential piece of CMB data is the
\COBE\ normalization of large-scale structure theories (see White \&
Scott (1996) for a discussion and list of references); however, with
the influx of data on smaller angular scales there are now extra
constraints which must be met.  For our purposes the strongest of
these is that our model must be able to produce sufficient
degree-scale power, which limits the amount to which we can tilt the
primordial spectrum to alleviate the large-scale structure problems.

We show the radiation anisotropy power spectrum for our models,
normalized to the \COBE\ four-year data, in Fig.~\ref{fig:cl}, along
with a standard CDM model for comparison. The angular power spectrum
is plotted in the usual way, with $\ell(\ell+1)C_\ell$ in
dimensionless units.  The index of the multipole expansion
$\ell\sim\theta^{-1}$, with $\theta$ in radians.  We have chosen not
to include any late reionization in these models, as the estimate of
the reionization redshift is very uncertain.  Including reionization
would reduce the power on scales $\ell\ga100$ by $\exp(-2\tau)$, where
estimates of $\tau$ are quoted in Table 1.

Notice that the increase in the baryon content has modulated the
height of the peaks in the spectrum.  The first and third peaks are
increased in amplitude while the second peak is decreased.  The tilt
of the spectrum has monotonically decreased the power at small angular
scales (high $\ell$).  This lack of power on smaller angular scales,
$\la15'$, is a robust feature of our models which should be testable
by the next generation of interferometers and array receivers.

Also shown in Fig.~\ref{fig:cl} is a selection of the current data on
CMB anisotropies.  From left to right the solid circles are: \COBE\
(Hinshaw et al. 1996; 3 points, 1 upper limit), SP94 \cite{Gunetal},
MAX \cite{Tanetal}, Python (Ruhl et al. 1996, 3-beam), MSAM (Inman et
al. 1996, 3-beam) and CAT (Scott et al. 1996, 2 points).  The error
bars are $\pm1\sigma$ while the horizontal lines indicate the range of
angular scales to which the experiment is primarily sensitive.  The
solid squares indicate the SK95 data \cite{Netetal}, with {\it no}
allowance made for the 14 per cent calibration uncertainty.  The upper
limits are the highest-$\ell$ point from \COBE\ at $\ell\simeq 30$,
then at large $\ell$ are White Dish \cite{Tucetal}, OVRO
\cite{Reaetal} and ATCA \cite{Subetal}, all quoted at 95 per cent
confidence.  For display purposes only, the points have been converted
from temperature to power (temperature squared) with error bars
$\sigma_{T^2} = 2T\sigma_T$.

These data have been selected from observations which have shown
repeatability and which probe angular scales near the first peak.  We
have not included any allowance in the error bars for the effects of
foreground subtraction.  The expected or measured size of the
foregrounds is discussed in the papers quoted above. Typically the
experiments have mapped regions for which foreground contamination is
known to be low and/or have used several frequencies to test for the
presence of foregrounds. The subtraction or correction for foregrounds
will generally weaken the lower limit on $n$.  Where experiments have
quoted two numbers from the same patch of sky we have chosen the
highest $\ell$, which provides the largest lever arm.  This is because
such points are correlated in an unknown way.  However, due to the
observing geometry and data reduction strategy, the SK95 points are
less than 20 per cent correlated \cite{Netetal}, except for a common
14 per cent calibration uncertainty, so we have included all 5 as
uncorrelated.  Corrections to this assumption should also permit a
lower $n$.

\begin{figure}
\begin{center}
\leavevmode
\epsfysize=6cm \epsfbox{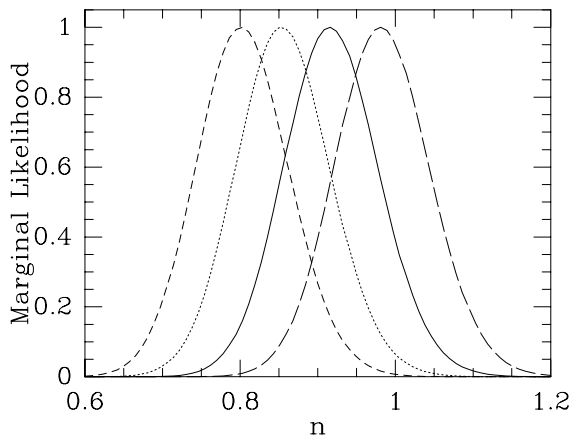}
\end{center}
\caption{The marginal likelihood versus~$n$, integrated over
normalization and SK95 calibration, for a fit to a selection of the
current CMB data.  We show likelihoods (from right to left) for
$\Ob=0.05$, $0.1$ (solid), $0.15$ and $0.2$. In all cases, $h$ is
fixed at 0.50.}
\label{fig:nfit}
\end{figure}

To test the constraining power of the CMB data on our theory, we have
computed the likelihood of fitting the data as a function of overall
normalization $\langle Q\rangle$, tilt $n$ and an additional
multiplicative factor for the SK95 points to represent the calibration
uncertainty.  We have not tried to include the upper limits or the CAT
points (which are correlated and have very large error bars). We show
the likelihood in Fig.~\ref{fig:nfit}, marginalized over the
normalization and SK95 calibration uncertainty. We have chosen a flat
prior for $\langle Q\rangle$, which is anyway well determined by the
\COBE\ data.  For the calibration uncertainty of SK95 we multiply by a
Gaussian centred on unity with width 14 per cent \cite{Netetal}. The
calibration uncertainty from the other experiments was added in
quadrature to the error bars on the points.

Our analysis here has not been very sophisticated, but it shows that
while the tilted model with 5 per cent baryons does indeed have
difficulty fitting the height of the peak \cite{WSSD}, models with 10
to 20 per cent baryons fare quite well.  It would be interesting to
perform a fit of these models to all of the data sets, rather than
just the band-powers, to see how well they fare in detail.  The best
fitting $n$ depends on the assumptions made about the tails of the
SK95 calibration uncertainty since the fit of any model to the data
becomes much better as the SK95 data are `lowered'.  The $\chi^2$ of
the best-fitting model decreases by $\sim10$ if the SK95 calibration
is decreased from $1$ to $0.82$ (consistent with the MSAM calibration)
for example.  With the SK95 calibration fixed at $1$ the best-fitting
model fits at about 95 per cent confidence level, this becomes 50 per
cent if the calibration is fixed at 0.82.  The first and second
moments of the marginal likelihoods from Fig.~\ref{fig:nfit} are
$n=0.98\pm0.06$, $0.92\pm0.06$, $0.86\pm0.06$ and $0.80\pm0.06$ for
$\Ob=0.05$, 0.10, 0.15 and 0.20 respectively.  If one includes the
possibility of reionization, then the $n = 0.8$ model with 10 per cent
baryons begins to become unlikely.  Including optical depth $\tau$, as
given by Table~\ref{tab:someoldnumbers}, is approximately equivalent
to shifting $n\to n-0.65\tau$.  Should observations further increase
the preferred peak height, then the model with 10 per cent baryons
will be disfavoured.  On the other hand, if the height were to
decrease, e.g.~if some of the anisotropy were non-cosmological, even
more tilt or reionization could be tolerated.

\section{Conclusions}

In this paper we have investigated the effects of assuming a baryon
fraction two or more times higher than 1993 estimates of BBN would
predict.  With $\Ob\ga0.1$ the baryons are becoming
dynamically relevant, and so the increase has a non-negligible impact
on the model of structure formation.  We have concentrated on cold
dark matter models with critical density, allowing $\Ob h^2$
to vary within the range $0.0125$ to $0.05$.

Inclusion of a higher baryon fraction damps small-scale power in these
models, since the growth in the baryons is suppressed by their
interactions with the (relativistic) photons.  More importantly, the
higher baryon fraction allows one to tilt the initial spectrum by a
greater amount before coming into conflict with the growing evidence
for enhanced power in the CMB on degree scales.

In order to facilitate more complete comparison of our model with data
and to encourage numerical experiments, we have chosen a `preferred'
model with $h=0.5$, $\Ob=0.12$ and $n=0.8$. This model is not the
absolute best-fit when considering large-scale structure data alone,
which would favour some combination of a slightly lower $h$, lower $n$
or higher $\Ob$, as seen in Figs.~\ref{fig:linh} and \ref{fig:linb}.
In selecting it, we have also paid heed to other evidence, such as
nucleosynthesis and direct measurement of the Hubble constant, which
suggest that one should avoid straying too far from the `standard'
values.  Clearly changes in these values of $\Ob$, $h$ and $n$ can
play off against each other.  Higher $\Ob$ allows less tilt, and vice
versa.  If reionization occurs early, or if future CMB data require
more power on degree scales, then this limits the tilt and thus the
highest $h$ which can be tolerated.

Several developments could falsify this model.  Firstly we require
that the Hubble constant be close to 50 km${\rm s}^{-1}$ ${\rm
Mpc}^{-1}$.  Should it be higher than 55 km${\rm s}^{-1}$ ${\rm
Mpc}^{-1}$, it would be very difficult to fit the current data while
retaining a critical matter density and pure CDM.  A higher Hubble
constant is in any case difficult for a critical-density universe
since it predicts an age of below $12\,$Gyr.  Of course, any firm
measurement of the total density $\Omega_0$ which obtained less than
one would also rule this model out.  In a similar vein, our model
predicts $\sigma_8$ to be in the interval from 0.5 to 0.8, and hence
that {\sl IRAS} galaxies are almost unbiased tracers of the mass. Both
results ought to be refutable using future analyses of, respectively,
velocity flows and redshift space distortions in galaxy surveys.
Furthermore, the standard theory of big bang nucleosynthesis is only
consistent with our assumed baryon fraction if the helium mass
fraction $Y_{{\rm P}}$ has been underestimated.  Should a firm upper
limit on $Y_{{\rm P}}$ around 24.5 per cent be made then this model
would require a violation of standard homogeneous nucleosynthesis by
the introduction of extra physics at $t\la3$ minutes.  We also require
the low deuterium determinations to be essentially correct, with the
implication of little destruction of primordial deuterium.

There are several specific predictions which the high $\Ob$
critical-density model makes.  Starting with the theoretically
cleanest of these, we predict that the radiation power spectrum will
have reduced power, compared to standard CDM, at $\ell \simeq 600$,
corresponding to angular scales $\simeq15'$.  This is seen clearly in
Figure~\ref{fig:cl}, and should be probed by the new generation of
interferometer experiments CBI, VCA and VSA, and ultimately by the
{\sl MAP} and {\sl COBRAS/SAMBA} satellites.  On large scales, the
shape of the matter power spectrum will differ from a scale-invariant
$\Gamma$-model, due to the tilt in the initial spectrum of
perturbations.  Since our model has critical density, objects do not
form as early as in open or cosmological constant dominated cosmologies, 
though the redshift of object formation is subject to large theoretical
uncertainties.  Finally, we predict a much higher fraction of dark
baryons, though we are unable to say what form they will take.
However, it is tempting to imagine that the halos of galaxies may have
around a 20 per cent contribution from low-mass stars and other
compact objects, as the recent results of the MACHO experiment
indicate \cite{MACHO2}.  We would certainly find it surprising if the
MACHO fraction in galaxies was significantly less than the baryonic
fraction in the Universe as a whole.

We briefly comment on the effect an increased baryon density would
have on other structure-formation models. In all cases, the effect of
the change to the CMB peak height is much more important than the
change to the matter power spectrum. Recall that for critical-density
CDM models, the main benefit is an increased CMB peak height, which is
necessary to compensate for the need to tilt in order to get the
galaxy correlations right. In low-density models, the spectral shape
is already corrected by the changed value of $\Omega_0 h$, and so tilt
is not required. However, observations only weakly constrain the tilt
in such models, and so changing the baryon density will have a fairly
neutral effect, offering neither advantages nor disadvantages --- it
simply moves the allowed region of parameters around a little.  For
high $\Ob/\Omega_0$, it may be possible to see oscillations in the
matter power spectrum from the baryon component, but at critical
density these are extremely small even in the 20 per cent model.  For
cold plus hot dark matter models on the other hand, a high baryon
density could be threatening, because one must have less tilt in those
models (the hot component having already adjusted the spectral shape)
and hence there is the danger of too high a peak in the CMB.  Early
reionization might then be necessary to reduce it to acceptable
levels.

Finally, let us end by noting that although critical-density CDM
models remain viable, the present observational situation allows only
a very limited range of parameters. If this paradigm is to be
believed, then already the value of $h$ is fixed at about the ten per
cent level near 0.50, and similarly the slope $n$ has to be around
$0.8 \pm 0.1$.

\section*{Acknowledgments}

MW is supported by the NSF and the DOE, PTPV by the PRAXIS XXI
programme of JNICT (Portugal), ARL by the Royal Society and DS by
NSERC (Canada). We thank Carlton Baugh, Marc Davis, Avery Meiksin,
Joel Primack, Joe Silk, Rachel Somerville and Helen Tadros for
discussions.  ARL thanks Columbia University, Stanford University and
Fermilab for hospitality while part of this work was carried out. PTPV
acknowledges the use of the Starlink computer system at the University
of Sussex.

\bsp

\end{document}